\documentclass[showpacs,twocolumn,prx,superscriptaddress,notitlepage]{revtex4-2}
\usepackage{qcircuit}
\usepackage[dvips]{graphicx}
\usepackage{amsmath,amssymb,amsthm,mathrsfs,amsfonts,dsfont}
\usepackage{subfigure, epsfig}
\usepackage{braket}
\usepackage{bm}
\usepackage{enumerate}
\usepackage{algorithm}
\usepackage{algpseudocode}
\usepackage{diagbox}
\usepackage{physics}
\usepackage{color}
\usepackage{multirow}
\usepackage[marginal]{footmisc}
\usepackage{comment}
\usepackage{tikz}
\usepackage[]{qcircuit}
\usetikzlibrary{arrows}
\usetikzlibrary{shapes,fadings,snakes}
\usetikzlibrary{decorations.pathmorphing,patterns}
\usetikzlibrary{calc}
\usetikzlibrary{positioning}
\usepackage[colorlinks = true]{hyperref}

\graphicspath{{./figure/}}

\newtheorem{observation}{Observation}

\newcommand{\comments}[1]{}

\begin{document}

\title{Entanglement Detection with Variational Quantum Interference: Theory and Experiment}

\author{Rui Zhang}
\altaffiliation{These authors contributed equally to this work.}
\affiliation{Hefei National Research Center for Physical Sciences at the Microscale and School of Physical Sciences, University of Science and Technology of China, Hefei 230026, China}
\affiliation{Shanghai Research Center for Quantum Science and CAS Center for Excellence in Quantum Information and Quantum Physics, University of Science and Technology of China, Shanghai 201315, China}

\author{Zhenhuan Liu}
\altaffiliation{These authors contributed equally to this work.}
\affiliation{Center for Quantum Information, Institute for Interdisciplinary Information Sciences, Tsinghua University, Beijing 100084, China}

\author{Chendi Yang}
\affiliation{Center for Quantum Information, Institute for Interdisciplinary Information Sciences, Tsinghua University, Beijing 100084, China}

\author{Yue-Yang Fei}
\affiliation{Hefei National Research Center for Physical Sciences at the Microscale and School of Physical Sciences, University of Science and Technology of China, Hefei 230026, China}
\affiliation{Shanghai Research Center for Quantum Science and CAS Center for Excellence in Quantum Information and Quantum Physics, University of Science and Technology of China, Shanghai 201315, China}
\affiliation{Hefei National Laboratory, University of Science and Technology of China, Hefei 230088, China}

\author{Xu-Fei Yin}
\affiliation{Hefei National Research Center for Physical Sciences at the Microscale and School of Physical Sciences, University of Science and Technology of China, Hefei 230026, China}
\affiliation{Shanghai Research Center for Quantum Science and CAS Center for Excellence in Quantum Information and Quantum Physics, University of Science and Technology of China, Shanghai 201315, China}
\affiliation{Hefei National Laboratory, University of Science and Technology of China, Hefei 230088, China}

\author{Yingqiu Mao}
\affiliation{Hefei National Research Center for Physical Sciences at the Microscale and School of Physical Sciences, University of Science and Technology of China, Hefei 230026, China}
\affiliation{Shanghai Research Center for Quantum Science and CAS Center for Excellence in Quantum Information and Quantum Physics, University of Science and Technology of China, Shanghai 201315, China}
\affiliation{Hefei National Laboratory, University of Science and Technology of China, Hefei 230088, China}

\author{Li Li}
\affiliation{Hefei National Research Center for Physical Sciences at the Microscale and School of Physical Sciences, University of Science and Technology of China, Hefei 230026, China}
\affiliation{Shanghai Research Center for Quantum Science and CAS Center for Excellence in Quantum Information and Quantum Physics, University of Science and Technology of China, Shanghai 201315, China}
\affiliation{Hefei National Laboratory, University of Science and Technology of China, Hefei 230088, China}

\author{Nai-Le Liu}
\affiliation{Hefei National Research Center for Physical Sciences at the Microscale and School of Physical Sciences, University of Science and Technology of China, Hefei 230026, China}
\affiliation{Shanghai Research Center for Quantum Science and CAS Center for Excellence in Quantum Information and Quantum Physics, University of Science and Technology of China, Shanghai 201315, China}
\affiliation{Hefei National Laboratory, University of Science and Technology of China, Hefei 230088, China}

\author{Yu-Ao Chen}
\affiliation{Hefei National Research Center for Physical Sciences at the Microscale and School of Physical Sciences, University of Science and Technology of China, Hefei 230026, China}
\affiliation{Shanghai Research Center for Quantum Science and CAS Center for Excellence in Quantum Information and Quantum Physics, University of Science and Technology of China, Shanghai 201315, China}
\affiliation{Hefei National Laboratory, University of Science and Technology of China, Hefei 230088, China}
\affiliation{New Cornerstone Science Laboratory, School of Emergent Technology, University of Science and Technology of China, Hefei 230026, China}

\author{Jian-Wei Pan}
\affiliation{Hefei National Research Center for Physical Sciences at the Microscale and School of Physical Sciences, University of Science and Technology of China, Hefei 230026, China}
\affiliation{Shanghai Research Center for Quantum Science and CAS Center for Excellence in Quantum Information and Quantum Physics, University of Science and Technology of China, Shanghai 201315, China}
\affiliation{Hefei National Laboratory, University of Science and Technology of China, Hefei 230088, China}

\begin{abstract}
Entanglement detection is a fundamental task in quantum information science, serving as a cornerstone for quantum benchmarking and foundational studies. 
With an increasing qubit number that can be effectively controlled, there is a pressing need for a scalable and robust detection protocol
which requires minimal resources while maintaining high detection capability. 
By integrating the Positive Partial Transposition criterion with variational quantum interference, we propose an entanglement detection protocol that requires moderate classical and quantum computation resources.
We numerically show that this protocol achieves a high detection capability with shallow quantum circuits, surpassing some widely-used entanglement detection methods.
The protocol also exhibits strong resilience to circuit noise, ensuring its applicability across different physical platforms. 
We further demonstrate the protocol experimentally on an eight-photon linear-optical platform, where it successfully detects the entanglement of a three-qubit mixed state that is inaccessible to conventional entanglement witnesses.
By combining quantum interference with classical optimization, our protocol provides a scalable and resource-efficient route toward practical entanglement detection.

\end{abstract}
\maketitle

\textbf{Introduction.}
Quantum entanglement, one of the distinguishing features of quantum physics~\cite{Horodecki2009entanglement}, provides advantages to a variety of quantum information processing tasks, including quantum computation~\cite{nielsen_chuang_2010, jozsa2003entanglement}, quantum cryptography~\cite{gisin2002cryptography,E91QKD}, and quantum metrology~\cite{giovannetti2004quantum,giovannetti2011advances}, as well as offers new perspectives for quantum many-body physics~\cite{amico2008entanglement,LAFLORENCIE2016quantum,abanin2019manybody}.
Therefore, the ability to generate large-scale entanglement has been widely utilized to showcase the superior capabilities of quantum devices~\cite{Haffner2005ion,Leibfried2005cat,monz2011fourteen,pan2012multiphoton,Islam2015purity,adam2016thermalization,wang2018sixphoton,omran2019cat,chao2019twenty,brydges2019probing,Cao2023super,shaw2023benchmarking}.
With the rapid development of quantum technologies enabling the manipulation of hundreds of qubits~\cite{Kim2023IBM,Bluvstein2024rydberg,jiang2025generation}, the pursuit of scalable, effective, and robust entanglement detection schemes constitutes a key objective in quantum information research.

In a typical entanglement detection experiment, researchers first prepare a target state and then experimentally verify its entanglement.
When the experimentally prepared state approximates the target state with sufficient fidelity, protocols including entanglement witnesses (EWs)~\cite{Haffner2005ion,Leibfried2005cat,monz2011fourteen,pan2012multiphoton,wang2018sixphoton,omran2019cat,chao2019twenty,Cao2023super} and Bell tests~\cite{Pan2000nonlocality,hyllus2005relation,moroder2013DI,Arnon2019DI,zhu2023entanglement} can be specially tailored to efficiently and effectively verify the entanglement.
However, the performance of these protocols is highly susceptible to unpredictable noise, leading to a rapid loss of detection capability under practical experimental conditions~\cite{liu2022fundamental,Miller2023graph}.
To address these limitations while eliminating dependence on a priori knowledge of the target state, researchers adopted the Positive Partial Transposition (PPT) criterion~\cite{peres1996ppt}, which exhibits reliable performance even for highly mixed states~\cite{aubrun2012PPTrandom,collins2016random}.
Nevertheless, the implementation of PPT criterion typically requires either complex joint operations~\cite{carteret2005peres} or classical post-processing of measurement data that scales exponentially with system size~\cite{elben2020mixed,zhou2020Single,Wang2022detecting}.
Alternative approaches incorporating variational quantum algorithms for parameter optimization, including machine learning-based methods~\cite{gray2018ml,lu2018ml,roik2021accuracy} and adversarial solvers~\cite{yin2022detection}, have emerged as promising directions. 
However, these methods demand either extensive computational resources for training data or the implementation of deep variational quantum circuits.

\begin{figure}[tbp]
\centering
\includegraphics[width=0.45\textwidth]{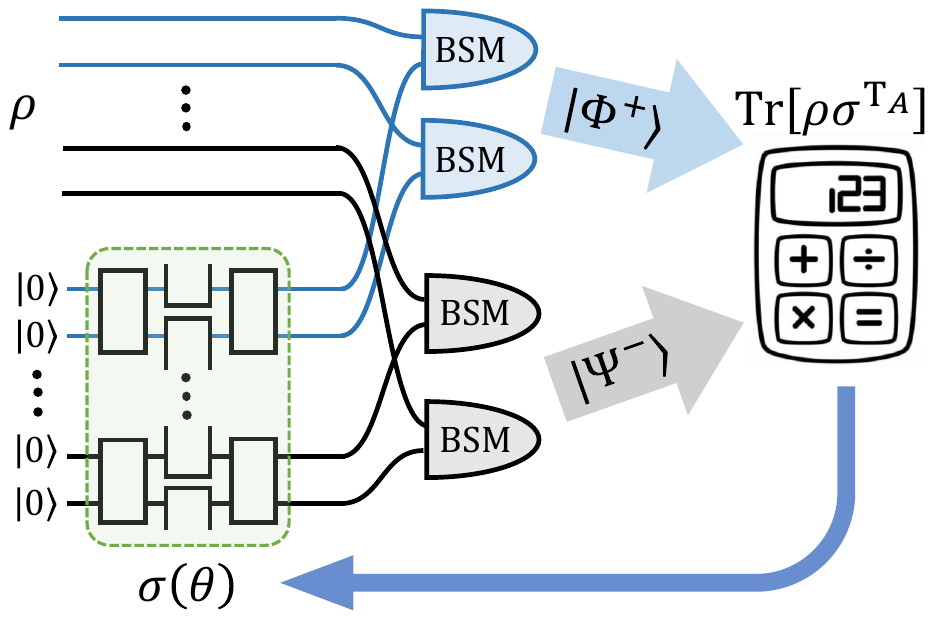}
\caption{Overview of iPPT protocol. 
For a target state $\rho$, the initial step involves preparing a reference state $\sigma(\theta)$ using the variational circuit (light green box). 
Subsequently, Bell state measurements (BSMs) are carried out on each corresponding qubit pair from the target state $\rho$ and the reference state $\sigma(\theta)$.
Lines in blue and black represent qubits of subsystems $A$ and $B$ of target and reference states.
BSM modules in blue and black collect data associated with the Bell states $\ket{\Phi^+}=(\ket{00}+\ket{11})/\sqrt{2}$ and $\ket{\Psi^-}=(\ket{01}-\ket{10})/\sqrt{2}$, respectively.
The collected data are post-processed to determine the entanglement property of the target state $\rho$, with a negative result of $\Tr(\rho\sigma^{\mathrm{T}_A})$ indicating the existence of entanglement.
}
\label{fig:main_figure}
\end{figure}

In this work, leveraging the detection capability of the PPT criterion and flexibility of variational quantum algorithms, we propose the \emph{interferometric PPT} (iPPT) protocol, as illustrated in Fig.~\ref{fig:main_figure}.
The iPPT protocol demonstrates advantages in both classical and quantum resource efficiency.
First, it avoids processing exponentially large classical datasets by eliminating the need for a priori knowledge of the reference state.
Second, inheriting the strong detection capability of PPT criterion, we numerically show that iPPT protocol exhibits superior detection capability compared to conventional methods such as fidelity~\cite{weilenmann2020faithful,gunhe2021faithful} and purity criteria~\cite{adam2016thermalization,brydges2019probing} with shallow reference state preparation circuits.
Notably, when a single layer of Bell state measurements (BSMs) is experimentally accessible, the iPPT protocol exhibits inherent robustness to noise in the reference state preparation circuit, a property derived from the PPT criterion. 
We experimentally implement the iPPT protocol to conduct a proof-of-principle entanglement detection in an eight-photon linear optical platform.
The experimental results highlight the effectiveness of our protocol in detecting entanglement in mixed states that conventional entanglement witness protocols fail to identify.

\textbf{Protocol.}
A bipartite state $\rho$ is entangled if and only if it cannot be decomposed in the form of
\begin{equation}
\rho=\sum_ip_i\rho_A^i\otimes\rho_B^i,
\end{equation}
where $\{p_i\}_i$ is a probability distribution and $\rho_A^i$ and $\rho_B^i$ are density matrices on subsystems $A$ and $B$.
For a quantum state $\rho$, the PPT criterion determines entanglement by examining the positivity of the partially transposed state $\rho^{\mathrm{T}_A}$, where $\mathrm{T}_A$ denotes the partial transposition with respect to subsystem $A$. 
The presence of negative eigenvalues in $\rho^{\mathrm{T}_A}$ certifies that $\rho$ is entangled.
Since partial transposition is not a physically implementable operation, it is essential to develop experimentally feasible entanglement detection protocols that retain the detection power of the PPT criterion while minimizing resource overhead.

Our approach is inspired by the EW-based PPT protocol~\cite{gunhe2009entanglement,rico2024poly}, which shows that verifying the PPT criterion is equivalent to identifying a reference state $\ket{\psi(\theta)}$ such that
\begin{equation}
\bra{\psi(\theta)}\rho^{\mathrm{T}_A}\ket{\psi(\theta)}
=\Tr\!\left[\rho\,\sigma(\theta)^{\mathrm{T}_A}\right]<0,
\end{equation}
where $\sigma(\theta)=\ketbra{\psi(\theta)}{\psi(\theta)}$.
In principle, one may search over a family of observables $\sigma(\theta)^{\mathrm{T}_A}$, which are entanglement witnesses, to detect a negative expectation value.
However, even when $\ket{\psi(\theta)}$ is restricted to shallow-circuit states, this protocol remains resource-intensive.
First, the observable $\sigma(\theta)^{\mathrm{T}_A}$ must be classically stored and processed, which involves handling an exponentially large matrix in the system size in order to determine an appropriate measurement scheme.
Second, measuring $\sigma(\theta)^{\mathrm{T}_A}$ requires rotating $\rho$ into the eigenbasis of this operator.
Although $\ket{\psi(\theta)}$ itself can be prepared by a shallow circuit, the unitary that diagonalizes $\sigma(\theta)^{\mathrm{T}_A}$ may require a significantly deeper circuit, as there is no straightforward correspondence between state-preparation circuits for $\ket{\psi(\theta)}$ and circuits that diagonalize $\sigma(\theta)^{\mathrm{T}_A}$.

To ease the difficulties, we propose the iPPT protocol, which is based on the following observation:
\begin{observation} \label{obs:main_obs}
Given two bipartite quantum states $\rho$ and $\sigma$, the following identity holds:
\begin{equation} \label{eq:entanglement_trace}
\Tr(\rho\sigma^{\mathrm{T}_A})
=\Tr\!\left[(\rho\otimes\sigma)(\Phi^+_A\otimes S_B)\right],
\end{equation}
where $\Phi^+_A=\sum_{i,j=0}^{2^{n_A}-1}\ketbra{ii}{jj}$ denotes the unnormalized maximally entangled state on subsystem $A$, and
$S_B=\sum_{i,j=0}^{2^{n_B}-1}\ketbra{ij}{ji}$ is the SWAP operator acting on subsystem $B$.
\end{observation}
When considering multi-qubit scenarios, $\Phi^+$ and $S$ all have tensor product structures, i.e., $\Phi^+=\bigotimes_{i}\Phi^+_i$ and $S=\bigotimes_{i}S_i$.
Here, $\Phi_i^+$ is the unnormalized Bell state $(\ket{00}+\ket{11})(\bra{00}+\bra{11})$ and the two-qubit SWAP operator can be written as $S_i=(\mathbb{I}-\Psi^-)$, where $\mathbb{I}$ is the identity operator and $\Psi^-=(\ket{01}-\ket{10})(\bra{01}-\bra{10})$ is another unnormalized Bell state.
Based on this observation, we can estimate the value of $\Tr(\rho\sigma^{\mathrm{T}_A})$ using BSMs performed between $\rho$ and $\sigma$, as shown by the circuit in Fig.~\ref{fig:main_figure}.
Specifically, labeling the measurement outcome of each BSM as $r_{i,j} \in \{0, 1, 2, 3\}$, which corresponds to the four Bell states $\{\ket{\Phi^+}$, $\ket{\Phi^-}$,  $\ket{\Psi^+}$, $\ket{\Psi^-}\}$, the unbiased estimator for $\Tr(\rho\sigma^{\mathrm{T}_A})$ is 
\begin{equation}
\Tr(\rho\sigma^{\mathrm{T}_A})=\prod_{i=0}^{(n_A-1)}\prod_{j=n_A}^{(n_A+n_B-1)}2\delta_{r_i,0}(1-2\delta_{r_j,3}),
\end{equation}
where $\delta_{i,j}=1$ when $i=j$ and $\delta_{i,j}=0$ otherwise.

The implementation of the iPPT protocol is illustrated in Fig.~\ref{fig:main_figure}.
One first prepares the reference state $\sigma(\theta)$ using a parameterized quantum circuit and then measures the observable $\Phi^+_A\otimes S_B$ via BSMs.
In contrast to the classical information processing required in the EW-based PPT protocol, the iPPT protocol enables the physical preparation of the parameterized reference state $\sigma(\theta)$, while the partial transposition operation is effectively encoded in the measured observable.
As a result, classical computation in the iPPT protocol is mainly used for processing BSM outcomes and updating the circuit parameters, which significantly reduces the classical overhead of the EW-based PPT protocol.
Moreover, the measurement procedure in the iPPT protocol requires only a single layer of two-qubit gates, making it readily applicable to a wide range of experimental platforms, such as linear optics~\cite{pan2012multiphoton} and cold atoms~\cite{Islam2015purity,adam2016thermalization}.
Furthermore, the iPPT protocol exhibits intrinsic robustness against errors in the reference-state preparation circuit, which may be substantially deeper than the BSM circuit.
This robustness stems from the fact that $\sigma(\theta)^{\mathrm{T}_A}$ remains a valid entanglement witness even when $\sigma(\theta)$ is a mixed state.
Therefore, as long as the BSMs can be implemented reliably, negative measurement outcomes can faithfully certify the presence of entanglement in the target state $\rho$.

\begin{figure}[htbp]
\centering
\includegraphics[width=0.45\textwidth]{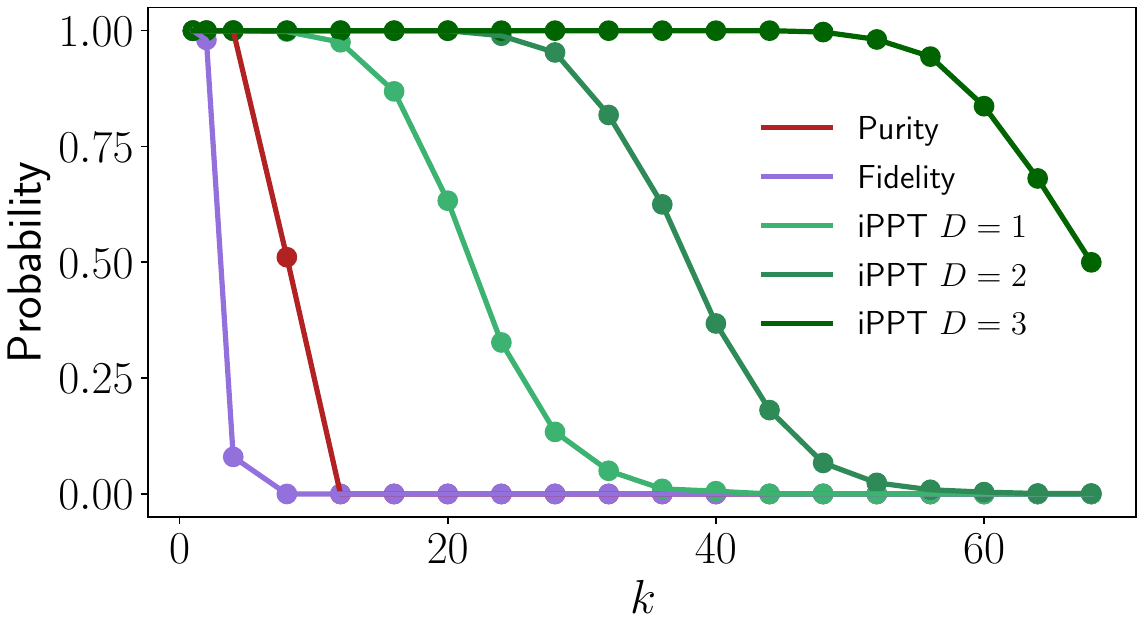}
\caption{
Entanglement detection capability of the iPPT protocol with various depth ($D$) of the reference state preparation circuits, in comparison with the purity and fidelity protocols.
For each point, we randomly sample 1000 six-qubit mixed states and verify their entanglement using different protocols.
The $y$-coordinate stands for the success probability for verifying the entanglement of the sampled six-qubit mixed state.
The larger values of $x$-coordinate term, $k$, correspond to the lower expected purity of the sampled mixed states. 
}
\label{fig:numberanalysis}
\end{figure}

\textbf{Numerical simulation.}
We demonstrate that the iPPT protocol inherits the strong detection power of the PPT criterion.
Using numerical simulations, we evaluate its detection performance for different reference-state circuit depths $D$ and benchmark it against purity~\cite{purity} and fidelity-based~\cite{fidelity} entanglement detection protocols.
These two protocols are chosen because they can also be implemented using Bell-state measurements, resulting in comparable experimental requirements, and have already been demonstrated on a wide range of physical platforms~\cite{Haffner2005ion,Leibfried2005cat,monz2011fourteen,pan2012multiphoton,omran2019cat,wang2018sixphoton,chao2019twenty,Cao2023super,Islam2015purity,adam2016thermalization,brydges2019probing,shaw2023benchmarking}.

To systematically assess the entanglement detection capabilities of these protocols on unknown states, we apply them to a set of six-qubit mixed states generated as follows.
We first sample a large number of random pure states in a $(2^6 \times k)$-dimensional Hilbert space and then trace out the $k$-dimensional subsystem to obtain six-qubit mixed states.
The parameter $k$ quantifies the degree of mixedness of the resulting states: larger values of $k$ correspond to lower purity and weaker entanglement.
As shown in Fig.~\ref{fig:numberanalysis}, the iPPT protocol significantly outperforms the purity and fidelity-based protocols, even when very shallow circuits are used, and retains a high detection capability for highly mixed states.
Further details of the numerical simulations are provided in Appendix~\ref{app:numerics}.

\begin{figure}[tbp!]
\centering
\includegraphics[width=0.95 \linewidth]{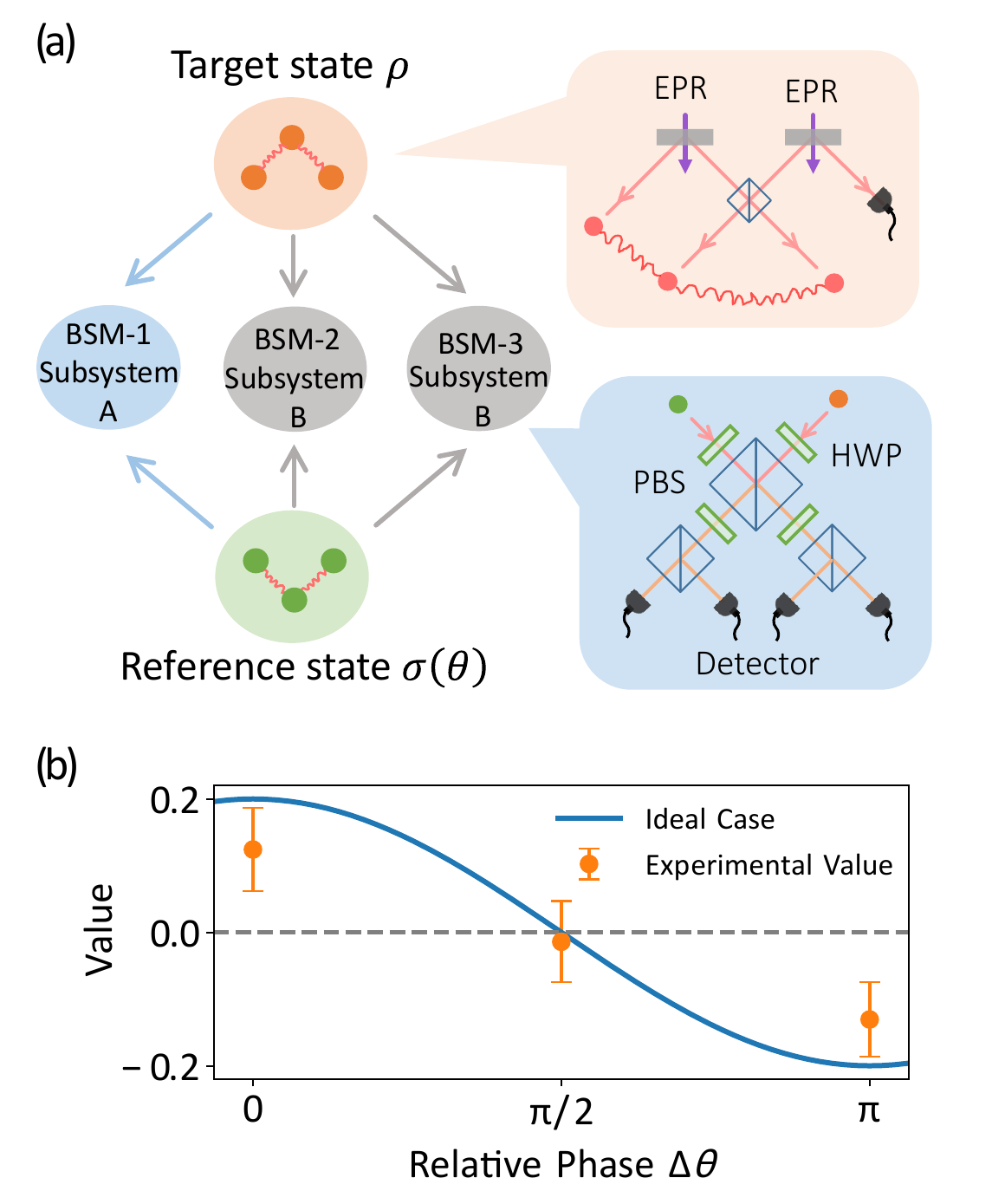}
\caption{Experimental design and results.
(a) The three-qubit target state $\rho$ and reference state $\sigma(\theta)$ are prepared with two pairs of entangled photons, respectively.
Three BSMs are performed between three pairs of photons from the reference and target state.
(b) Ideal and experimental value of $\Tr(\rho\sigma^{\mathrm{T}_A})$ with respect to the variable phase parameter $\Delta \theta$.
}
\label{fig:protocol_exp}
\end{figure}

\begin{figure*}[hbtp!]
\centering
\includegraphics[width= 1\linewidth]{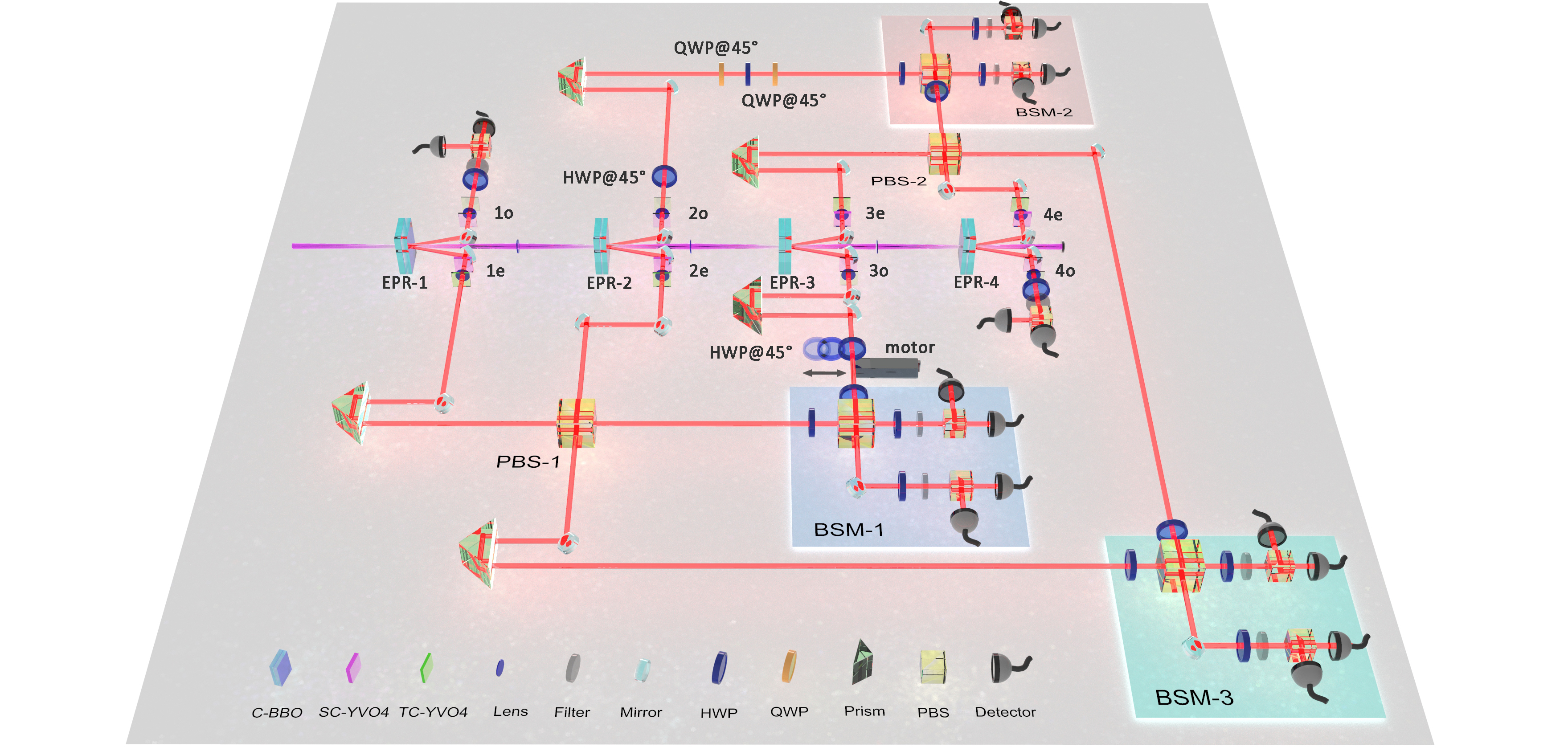}
\caption{Schematic of experimental setup for iPPT protocol.
The pulsed ultraviolet pump laser successively passes through four separated C-BBO crystals from left to right and generate four EPR entangled photon pairs, $\ket{\mathrm{EPR}}=(\ket{HH}+\ket{VV})/\sqrt{2}$.
The first two EPR sources, EPR-1 and EPR-2, are used to prepare the reference state.
The last two EPR sources, EPR-3 and EPR-4, are used to prepare the to-be-verified target state. 
The mixing of each component state of the target state is achieved by a one-dimensional linear motor (``motor" in schematic) that randomly introducing the polarization flipping.
The three pairs of photons from reference state and target state are introduced to three BSM device, which are composed of PBSs, HWPs, and single-photon detectors.
SC-YVO4 and TC-YVO4 are YVO$_4$ crystals used for spatial compensation (SC) and temporal compensation (TC) of each EPR source.
}
\label{fig:setup_figure}
\end{figure*}

\textbf{Experimental demonstration.}
After numerically demonstrating the detection capability, we devise a proof-of-principle experiment in linear optical platform to show the practicality of iPPT protocol.
We consider the three-qubit bipartite state
\begin{equation}\label{eq:target_state}
\rho=0.5\rho_1 + 0.5\rho_2,
\end{equation}
where $\rho_\mathrm{1}$ and $\rho_\mathrm{2}$ are Greenberger–Horne–Zeilinger (GHZ) states with phase-flipping error, 
\begin{equation}\label{eq:target_state_component}
\begin{aligned}
\rho_1=&F^{+}_1(\ket{000}+\ket{111})(\bra{000}+\bra{111})\\
+&F^{-}_1(\ket{000}-\ket{111})(\bra{000}-\bra{111}),\\
\rho_2=&F^{+}_2(\ket{100}+\ket{011})(\bra{100}+\bra{011})\\
+&F^{-}_2(\ket{100}-\ket{011})(\bra{100}-\bra{011}).
\end{aligned}
\end{equation}
By setting four coefficients to be $F^{+}_1=F^{+}_2=0.9$ and $F^{-}_1=F^{-}_2=0.1$, the proportions of all four main component states of $\rho$ are less than 0.5.
We show in Appendix~\ref{app:EW} that its entanglement cannot be identified with conventional fidelity-based entanglement witness protocols~\cite{gunhe2009entanglement}. 
With the prior knowledge of the target state $\rho$, we can customize the parameterized reference state $\sigma(\Delta \theta)=\ket{\psi(\Delta \theta)}\bra{\psi(\Delta \theta)}$ with
\begin{equation}\label{eq:reference_state}
\ket{\psi(\Delta \theta)}=\frac{1}{\sqrt{2}}(\ket{010}+ \mathrm{e}^{i \Delta \theta} \ket{101}),
\end{equation}
to evaluate its entanglement, where $\Delta \theta$ is the variable phase parameter. 
By assigning the first qubit as the subsystem $A$ and the other two qubits as the subsystem $B$, as shown in Fig.~\ref{fig:protocol_exp}(a), the value of $\Tr\left[\rho\sigma(\Delta \theta)^{\mathrm{T}_A}\right]$ with respect to the phase parameter $\Delta \theta$ can be calculated as the blue solid curve in Fig.~\ref{fig:protocol_exp}(b).
When $\Delta \theta$ is in the interval $(\pi/2,\pi)$, the value of $\Tr\left[\rho\sigma(\Delta \theta)^{\mathrm{T}_A}\right]$ is negative, and the entanglement in the mixed target state $\rho$ can be clearly identified.

We perform the experimental demonstration of iPPT entanglement detection protocol in an eight-photon entanglement platform as depicted in Fig.~\ref{fig:setup_figure}. 
In the setup, a pulsed ultraviolet laser beam with a central wavelength of 390~$\mathrm{nm}$ is successively focused on four sandwich-like $\beta$-barium borate (C-BBO) crystal combinations. 
Through the type-II spontaneous parametric down-conversion (SPDC) process, each C-BBO generates a pair of Einstein–Podolsky–Rosen (EPR) entangled photons, $\ket{\mathrm{EPR}}=(\ket{HH}+\ket{VV})/\sqrt{2}$, where $\ket{H}$ and $\ket{V}$ denote horizontal and vertical polarization encoding states $\ket{0}$ and $\ket{1}$, respectively. 
We utilize the first two EPR sources (EPR-1 and EPR-2, photons 1e-2o-2e) to prepare the parameterized reference state $\sigma(\Delta \theta)$, while the last two sources (EPR-3 and EPR-4, photons 3e-3o-4e) are dedicated to preparing the target state $\rho_\mathrm{exp}$ in the form of Eq.~\eqref{eq:target_state}.
Following state preparation, the relevant photons from the reference and target states are directed into three polarizing beam splitters (PBSs) to execute BSMs.
Each BSM module, consisting of half-wave plates (HWPs), PBSs, and single-photon detectors, is capable of distinguishing two out of the four Bell states in a single trial~\cite{pan1998BSM}.
Specifically, similar with the illustration of Fig.~\ref{fig:protocol_exp}(a), BSM-1 is applied to subsystem A to perform the measurement of the Bell state $\ket{\Phi^+}$, whereas BSM-2 and BSM-3 are applied to subsystem B perform the measurement of the Bell state $\ket{\Psi^-}$. 

In our implementation, the pump laser power is set to $400~\mathrm{mW}$, and band-pass filters with a full width at half maximum (FWHM) of $4~\mathrm{nm}$ are applied to the down-converted photons to ensure spectral indistinguishability. 
With these settings, the target state is prepared in the form given by Eqs.~\eqref{eq:target_state} and ~\eqref{eq:target_state_component}, achieving a fidelity of 0.977 with respect to the ideal state.
The measured proportions of the four main component states are $\frac{F^{+}_1}{2}=0.418$, $\frac{F^{-}_1}{2}=0.060$,  $\frac{F^{+}_2}{2}=0.448$, and $\frac{F^{-}_2}{2}=0.035$.
Crucially, since all these coefficients are less than 0.5, the entanglement of the prepared target state cannot be detected using conventional fidelity-based EW methods~\cite{gunhe2009entanglement}. 
To characterize the measurement apparatus, the Hong-Ou-Mandel-type two-photon interference visibilities at the three PBSs are measured to be $81.9\%$, $83.6\%$, and $84.9\%$, respectively.

We execute the iPPT protocol to verify the entanglement of the prepared three-qubit mixed target state $\rho_\mathrm{exp}$.
To illustrate the variational parameter optimization, we choose three representative phase settings $\Delta \theta$ and estimate the corresponding values of $\Tr\left[\rho\sigma(\Delta \theta)^{\mathrm{T}_A}\right]$ from the BSM outcomes, shown as orange dots in Fig.~\ref{fig:protocol_exp}(b).
The experimental data agree well with the theoretical prediction (blue solid curve), apart from a slight systematic reduction in amplitude, which we attribute to non-ideal two-photon interference visibility in the BSM setup.
At the optimal parameter $\Delta\theta=\pi$, we obtain $\Tr\left[\rho\sigma(\Delta \theta)^{\mathrm{T}_A}\right] = -0.131 \pm 0.056$.
Although the statistical confidence is limited by the low count rates of eight-photon coincidence measurements, the consistent trend across the parameter space supports the validity of this negative result.
Therefore, despite statistical fluctuations, our data provide strong evidence for entanglement in the mixed target state $\rho_\mathrm{exp}$ and demonstrate the robustness of the iPPT protocol against realistic physical noise, as long as interference visibility is not completely lost.

\textbf{Discussion.}
Entanglement detection is widely recognized as a challenging task~\cite{liu2022fundamental}.
Although the iPPT protocol does not resolve all challenges associated with entanglement detection, it introduces a practical framework that improves classical efficiency, robustness, and scalability.
Moreover, the iPPT protocol provides a flexible mechanism to trade detection capability for quantum resources.
As shown in Fig.~\ref{fig:numberanalysis}, the detection power of the iPPT protocol increases with the depth of the reference-state preparation circuit.
However, due to the trade-off between expressibility and trainability in variational quantum algorithms~\cite{McClean2018BP,holmes2022connecting,yu2023expressibility}, increasing circuit depth also makes parameter optimization more challenging.
Therefore, in practical implementations, it is crucial to exploit prior knowledge of the target state $\rho$ to design more efficient ansatz structures, such as the one in Eq.~\eqref{eq:reference_state}.

Beyond the PPT criterion, Observation~\ref{obs:main_obs} can be generalized to an arbitrary positive-map criterion of the form
\begin{equation}\label{eq:general_positive_map}
\Tr[\mathcal{N}_A\otimes\mathcal{I}_B(\rho)\psi]
=\Tr[(\rho\otimes\psi)(W_A\otimes S_B)],
\end{equation}
where $\mathcal{N}$ is a positive map and $W$ is an EW related with $\mathcal{N}$.
Since positive-map criteria typically possess exponentially stronger detection power than EWs~\cite{liu2022fundamental}, Eq.~\eqref{eq:general_positive_map} offers a systematic approach to enhancing the detection capability of EWs by leveraging the same underlying logic as the iPPT protocol.

\textbf{Acknowledgements.}
We appreciate the helpful discussions with Xiongfeng Ma, Otfried G\"uhne, Satoya Imai, Andreas Elben, Daniel Miller, You Zhou, Hong-Ye Hu, Zhenyu Cai, Feihu Xu,  and Ping Xu. 
This work was supported by the National Natural Science Foundation of China (Grants No.~11975222, No.~11874340, No.~12174216), Shanghai Municipal Science and Technology Major Project (Grant No.~2019SHZDZX01), Chinese Academy of Sciences and the Shanghai Science and Technology Development Funds (Grant No.~18JC1414700), the Innovation Program for Quantum Science and Technology-National Science and Technology Major Project (Grant No.~2021ZD0301901, No.~2021ZD0300804), the Basic Science Center Project of NSFC (Grant No. 12488301), and the New Cornerstone Science Foundation.
Xu-Fei Yin was support from the China Postdoctoral Science Foundation (Grant No.~2023M733418).
The numerical simulations are mainly performed by using MindSpore Quantum~\cite{xu2024mindspore} with updated code available at \href{https://gitee.com/mindspore/mindquantum/tree/research/paper_with_code/entanglement_detection_with_variational_quantum_interference}{Gitee}.

\comments{
\begin{enumerate}
\item Can we use this technique to solve other semi-definite programming problems?
\item Can we derive some analytical results about the detection capability of our method with a fixed circuit depth?
\item Can we add more restrictions on the reference state, like the Schmidt rank?
\end{enumerate}
}

\bibliography{iPPTref}

\begin{thebibliography}{55}
\bibitem [{pur()}]{purity}%
  \BibitemOpen
  \href@noop {} {}\bibinfo {note} {The purity criterion states that when $\Tr(\rho_A^2)<\Tr(\rho^2)$ with $\rho_A$ being the reduced density matrix of $\rho$, then $\rho$ is entangled.}\BibitemShut {Stop}%
\bibitem [{fid()}]{fidelity}%
  \BibitemOpen
  \href@noop {} {}\bibinfo {note} {The fidelity criterion states that when $\bra{\psi}\rho\ket{\psi}>\alpha$ with $\alpha$ being the square of the maximal Schimdt value of $\ket{\psi}$, then $\rho$ is entangled.}\BibitemShut {Stop}%
\end{thebibliography}

\appendix

\onecolumngrid
\newpage

\section{Theoretical Analysis}
\subsection{Generalization to arbitrary positive maps}
Using the Choi matrix representation, the action of a map can be represented as
\begin{equation}
\mathcal{N}(\rho)=\Tr\left[\Lambda_{\mathcal{N}}\left(\rho^{\mathrm{T}}\otimes\mathbb{I}\right)\right]
=\begin{tabular}{c}
\includegraphics[scale=0.35]{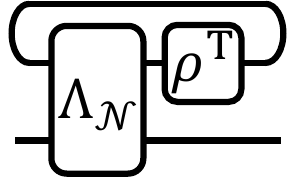}
\end{tabular}
=\begin{tabular}{c}
\includegraphics[scale=0.35]{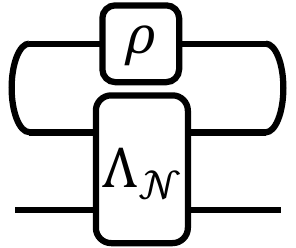}
\end{tabular},
\end{equation}
where $\Lambda_{\mathcal{N}}$ is the Choi matrix of map $\mathcal{N}$, legs represent indices of matrices, and legs connection stands for the indices contraction.
With this tool, the inner product can be represented as
\begin{equation}
\Tr\left[\mathcal{N}_A\otimes\mathcal{I}_B(\rho)\psi\right]
=\begin{tabular}{c}
\includegraphics[scale=0.35]{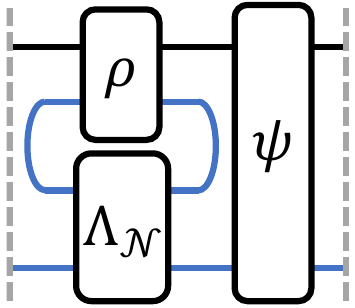}
\end{tabular}
=\begin{tabular}{c}
\includegraphics[scale=0.35]{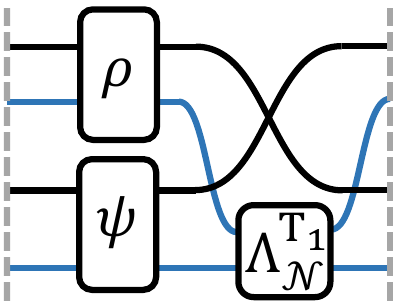}
\end{tabular},
\end{equation}
where the grey dashed lines represent trace functions, blue and black legs represent indices of subsystem $A$ and $B$, respectively.
As is known, the Choi matrix of a positive map, together with its partial transposition, are valid entanglement witnesses~\cite{HORODECKI1996separability}.
Therefore, we can prove that
\begin{equation}
\Tr\left[\mathcal{N}_A\otimes\mathcal{I}_B(\rho)\psi\right]=\Tr\left[\left(\rho\otimes\psi\right)\left(W_A\otimes S_B\right)\right].
\end{equation}
Assuming $\mathcal{N}$ to be the transposition map, we have the main result of this work, $\Tr\left[\rho^{\mathrm{T}_A}\psi\right]=\Tr\left[\left(\rho\otimes\psi\right)\left(\Phi^+_A\otimes S_B\right)\right]$.

\subsection{Details of numerical experiments}\label{app:numerics}
In the main context, we discussed the fidelity-based entanglement witness protocol. 
Specifically, this protocol is based on the entanglement witness
\begin{equation} \label{EW}
W=\alpha\mathbb{I}-\ketbra{\psi}{\psi},
\end{equation}
where $\mathbb{I}$ is the identity operator and $\alpha$ is the square of the maximal Schimdt number of $\ket{\psi}$.
The verification of this protocol is equivalent with verifying if the fidelity between the target state $\rho$ and $\ket{\psi}$ is larger than $\alpha$ or not.
Another entanglement detection protocol employed in the main context, the purity-based protocol, states that, when
\begin{equation}
\Tr(\rho_A^2)-\Tr(\rho^2)<0,
\end{equation}
where $\rho_A$ is the reduced density matrix of $\rho$, the state $\rho$ is bipartite entangled.

As stated in the main context, the fidelity-based entanglement witness and the purity entanglement detection protocol can be verified using the same circuit structure as the iPPT criterion.
For purity protocol, this is because
\begin{equation}
\Tr(\rho_A^2)-\Tr(\rho_{AB}^2)=\Tr\left[(\rho\otimes\rho)(S_A\otimes\mathbb{I}_B-S_{AB})\right],
\end{equation}
where the expectation value of $S_A\otimes\mathbb{I}_B-S_{AB}=S_A\otimes(\mathbb{I}_B-S_B)$ can also be estimated using BSM.
Thus, by changing the reference state $\ket{\psi}$ into the same state as the target state, one can use the same circuit structure as iPPT to estimate the purity criterion.
For the fidelity protocol, our conclusion is made according to the observation
\begin{equation}
\Tr(W\rho)=\Tr\left[(\rho\otimes\psi)(\alpha\mathbb{I}-S_{AB})\right],
\end{equation}
where $\alpha\mathbb{I}-S_{AB}$ can also be estimated using BSM.
However, different with iPPT and the purity protocol, the value of $\alpha$ need to be classically calculated according to the classical description of $\ket{\psi}$, which increases its classical computational resources consumption.
Due to these reasons, we compare the iPPT protocol with these two entanglement detection protocols.

\begin{figure}[htbp]
\centering
\includegraphics[width=0.35\textwidth]{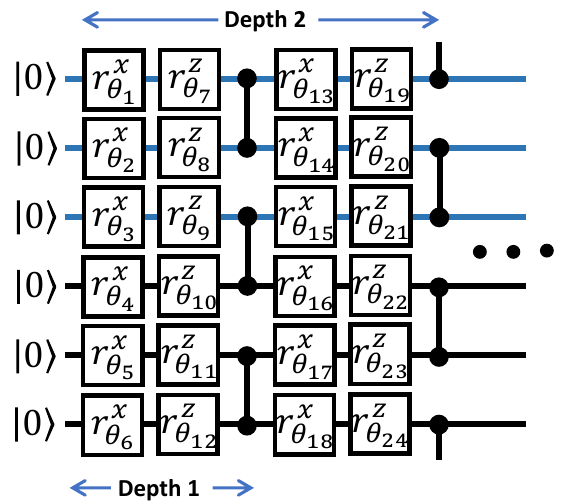}
\caption{Illustration of the variational circuit for preparing the reference state. 
}
\label{fig:circuit}
\end{figure}

In the numerical experiments Fig.~\ref{fig:numberanalysis}, we use the variational circuit shown in Fig.~\ref{fig:circuit} to search for the reference state.
A total of six qubits is divided into two groups, acting as subsystems $A$ and $B$.
A single layer of the circuit is composed of two layers of parameterized single-qubit rotation gates, $r_\theta^x=e^{i\theta X}$ and $r_\theta^z=e^{i\theta Z}$, followed by a layer of controlled-not gates.
We say a circuit is of depth-$k$ if it contains $k$ layers.
The power of iPPT protocol can be also reflected from this definition as a single layer of controlled-not gates already achieves good detection capability.

To calculate the detection probabilities, for each value of $k$, we sample $1000$ different $2^6\times k$ dimensional Haar random pure states.
For each pure state, we trace out the $k$-dimensional system and leave the $2^6$ dimensional mixed state.
Then, we detect these $1000$ mixed states using iPPT protocol with different reference state circuit depth, fidelity-based EW criterion, and the purity criterion.
The probabilities are calculated using the ratio of the successfully detected states.

To compare with the fidelity-based entanglement witness protocol, we variationally search over all pure states that can be prepared via the depth-3 circuit $\ket{\psi}$, calculate the fidelity between $\ket{\psi}$ and the target state $\rho$, and compare it with the square of the largest Schimdt number of $\ket{\psi}$.

\subsection{Fidelity-based entanglement witness}\label{app:EW}

As introduced in Eq.~\eqref{EW}, entanglement in a quantum state $\rho$ can be effectively detected with the implementation of an entanglement witness operator $\mathit{W} = \alpha \mathbb{I}-\ket{\psi}\bra{\psi}$, where a negative expectation value, $\left\langle W \right\rangle = \Tr(\rho \mathit{W})<0$, indicates the presence of entanglement. 
The main idea of the construction of above EW operator is measuring the distance between the prepared state and the target state $\ket{\psi}$, such as the state fidelity, to determine the entanglement property of state $\rho$.
The construction of the above entanglement witness operator is centered around evaluating the state fidelity, i.e., measuring the proximity between the prepared state $\rho$ and the pure entangled target state $\ket{\psi}$, so as to assess the entanglement characteristics of the state $\rho$. 
Benefiting from its relatively low measurement requirement and resource consumption, EW is effective for some simple quantum states in the experiment. 
Nevertheless, constructing an entanglement witness necessitates prior knowledge about the prepared state $\rho$, such as the specific expression of the target state $\ket{\psi}$. 

In the main text, the mixed state to be measured is represented as
\begin{equation}\label{eq:target_state_ideal}
\begin{aligned}
\rho_{\mathrm{ideal}}=&0.45(\ket{000}+\ket{111})(\bra{000}+\bra{111})
+0.05(\ket{000}-\ket{111})(\bra{000}-\bra{111})\\
+&0.45(\ket{100}+\ket{011})(\bra{100}+\bra{011})
+0.05(\ket{100}-\ket{011})(\bra{100}-\bra{011}).
\end{aligned}
\end{equation}
As indicated by its explicit expression, 
the closest pure entangled states to it
are two GHZ-type state: $\ket{\text{GHZ}}_1=(\ket{000}+\ket{111})/\sqrt{2}$ and $\ket{\text{GHZ}}_2=(\ket{100}+\ket{011})/\sqrt{2}$.
Neither of these two pure states is adequate for constructing an effective EW operator since the proportions of both are smaller than the largest Schmidt number denoted by $\alpha = 0.5$. 
Upon calculation, it is evident that the corresponding expectation value, $\left\langle W\right \rangle = 0.05 >0$, indicating the inability to identify the entanglement of the quantum state.
However, this does not ensure that there are no other three-qubit pure states that could potentially serve as a basis for constructing an effective EW operator to detect entanglement in the mixed state $\rho_{\mathrm{ideal}}$.
To eliminate the possibility of a pure state that can be used to construct an effective EW operator, we systematically examine all three-qubit pure states within the general expression of $\ket{\phi}=c_0 \ket{000}+c_1 \ket{001}+...+c_7 \ket{111}$, where the coefficients $c_i$ are complex numbers satisfying the normalization condition $\sum_{i=0}^{7}\left| c_{i} \right|^{2}=1$.
Upon systematically analyzing all three-qubit pure states numerically, we determined that the minimum EW value is $\left\langle W\right\rangle_{\mathrm{min}}=0.05>0$. 
This observation underscores the incapacity of the fidelity-based conventional EW to assess entanglement in the mixed state $\rho_{\mathrm{ideal}}$. 

\section{Experimental Details of State Preparation}\label{app:exp}

\subsection{preparation of the reference state $\ket{\psi(\Delta \theta)}$} 
In our experiment, EPR photon pairs from the first two entanglement sources are ``fused'' into a four-photon GHZ state by interfering photons $1$e and $2$e on one polarizing beam splitter (PBS), PBS-1. 
On this basis, the reference state, $\ket{\psi(\Delta \theta)}=(\ket{HVH}+\mathrm{e}^{i\Delta \theta}\ket{VHV})/{\sqrt{2}}$, can be prepared on photons $1$e-$2$o-$2$e by triggering the heralding photon $1$o into the diagonal state $\ket{+}=(\ket{H}+\ket{V})/\sqrt{2}$, rotating polarization state of photon $2$o with a half-wavelength plate (HWP), and introducing relative phase $\Delta \theta$ with the wave plate combination.

\subsection{preparation of the target state $\rho_{\mathrm{exp}}$}
\begin{figure*}[htbp]
\centering	\includegraphics[width=0.9 \linewidth]{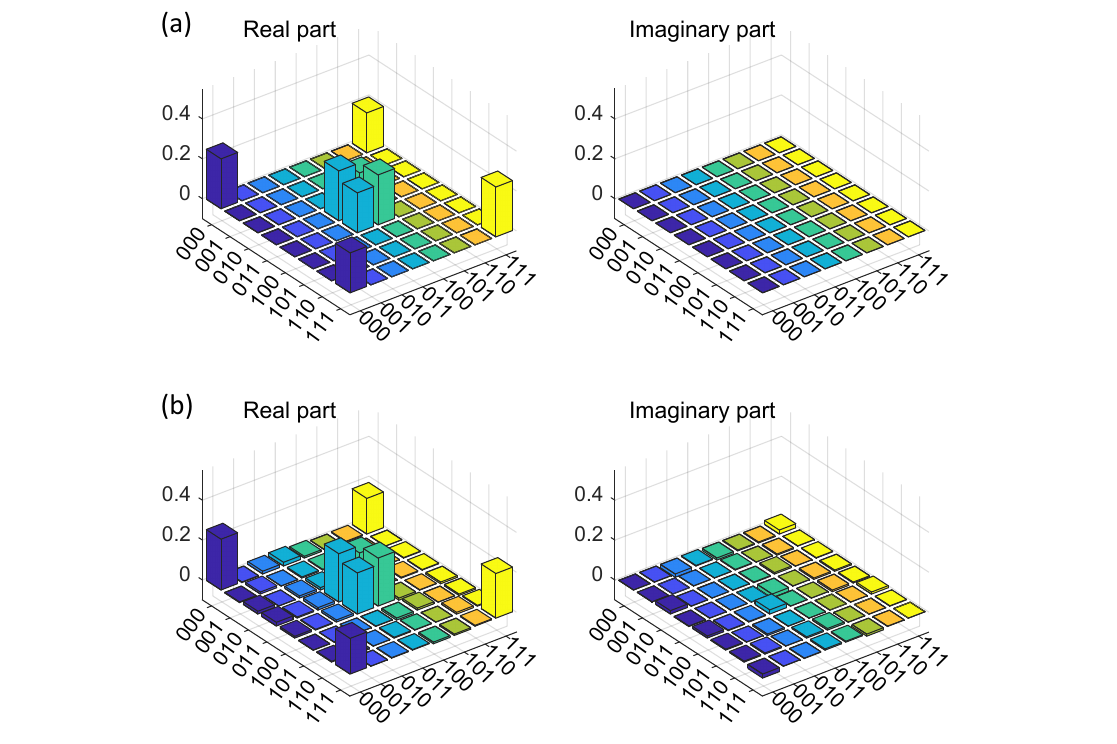}
\caption{ 
(a) The density matrix of the ideal target state.
(b) The density matrix of experimentally reconstructed target state.
}
\label{Fig:tomo}
\end{figure*}

As for preparation of the to-be-verified target state $\rho_{\mathrm{exp}}$, the EPR pairs from the last two entanglement sources are first introduced to prepare the three-photon GHZ state, $\ket{\mathrm{GHZ}_3}=(\ket{HHH}+\ket{VVV})/{\sqrt{2}}$, by interfering photons $3$e and $4$e on PBS-2 and triggering the heralding photon $4$o into the diagonal state $\ket{+}$. 
Then, polarization flipping is randomly introduced to the photon $3$o with an HWP@45$^{\circ}$ driven by a linear stepping motor, transforming $\ket{\mathrm{GHZ}_3}$ into a mixed state, 
\begin{equation}\label{eq:mixed_GHZ}
\begin{aligned}
\rho_{\mathrm{GHZ}}=&\frac{1}{2}(\ket{HHH}+\ket{VVV})(\bra{HHH}+\bra{VVV})\\
+&\frac{1}{2}(\ket{VHH}+\ket{HVV})(\bra{VHH}+\bra{HVV}).
\end{aligned}
\end{equation}
Due to the imperfect two-photon interference on PBS-2 that mainly induces phase flip error, the to-be-verified target state in form of Eq.~\eqref{eq:target_state} can be actually prepared,
\begin{equation}\label{eq:target_state_exp_app}
\begin{aligned}
\rho_{\mathrm{exp}}=&\frac{F^{+}_1}{2}(\ket{HHH}+\ket{VVV})(\bra{HHH}+\bra{VVV})
+\frac{F^{-}_1}{2}(\ket{HHH}-\ket{VVV})(\bra{HHH}-\bra{VVV})\\
+&\frac{F^{+}_2}{2}(\ket{VHH}+\ket{HVV})(\bra{VHH}+\bra{HVV})
+\frac{F^{-}_2}{2}(\ket{VHH}-\ket{HVV})(\bra{VHH}-\bra{HVV})\\
+&\frac{F^{\prime} }{2} \rho^{\prime}_{\mathrm{GHZ}}. 
\end{aligned}
\end{equation}
Coefficients $F^{\pm}_{i}$ are mainly determined by the two-photon interference visibility of the PBS-2.  
The component state $\rho^{\prime}_{\mathrm{GHZ}}$ represents other polarization-flipping noisy states generated during state preparation process, whose proportion $F^{\prime}$ is generally smaller than $0.05$.
The four GHZ states, $(\ket{HHH} \pm \ket{VVV})/\sqrt{2}$ and $(\ket{VHH} \pm \ket{HVV})/\sqrt{2}$, are the main component states of the target state.
The component state $\rho^{\prime}_{\mathrm{GHZ}}$ represents all other six noisy states, which are generated during the whole state preparation process.
We perform the quantum state tomography measurement on the prepared target state and reconstruct its density matrix as shown in Fig.~\ref{Fig:tomo}(b).
The experimentally prepared target state is very close to the ideal case,
\begin{equation}\label{eq:target_state_ideal}
\begin{aligned}
\rho_{\mathrm{ideal}}=&0.45(\ket{HHH}+\ket{VVV})(\bra{HHH}+\bra{VVV})
+0.05(\ket{HHH}-\ket{VVV})(\bra{HHH}-\bra{VVV})\\
+&0.45(\ket{VHH}+\ket{HVV})(\bra{VHH}+\bra{HVV})
+0.05(\ket{VHH}-\ket{HVV})(\bra{VHH}-\bra{HVV}),
\end{aligned}
\end{equation}
with the fidelity is calculated to be $0.977$.

\end{document}